\documentstyle[12pt]{article}

\def\k{{\rm {\bf k}}}
\def\p{{\rm {\bf p}}}
\def\q{{\rm {\bf q}}}
\def\Re{{\rm Re}}

\def\exp{{\rm exp}}
\def\g{\mbox{\boldmath$\gamma$}}

\begin{document}

\vspace{1.5cm}

\phantom{...}

\vspace{2cm}

\begin{center}
{\bf PHOTON AND ELECTRON SPECTRA IN HOT AND DENSE QED}

\vspace{1.5cm}

{\bf O.K.Kalashnikov}
\footnote{ E-mail address: kalash@td.lpi.ac.ru}

Theoretical Physics Division,

P.N.Lebedev Physical Institute,

Russian Academy of Sciences,

117924 Moscow, Russia.

\vspace{2.5cm}

{\bf Abstract}
\end{center}

Photon and electron spectra in hot and dense QED are found in the high
temperature limit  for all $|\q|$  using the Feynman gauge and the
one-loop self-energy. All spectra are split by the medium and their
branches develop the gap (the dynamical mass) at zero momentum. The
photon spectrum has two branches (longitudinal and transverse) with the
common mass; but electron spectrum is split on four branches which are
well-separated for any $|\q|$ including their $|\q|=0$ limits (their
effective masses). These masses and the photon thermal mass are
calculated explicitly and the different limits of spectrum branches
are established in detail. The gauge invariance of the high-temperature
spectra is briefly discussed.

\newpage

\section{Introduction}
At present much attention is paid to studying the collective
excitation spectra in hot and dense medium, using in the first rate the
standard QCD medium or GUT models. In any medium collective excitations
arise due to the multi-interaction with the heat bath and one another,
and unlike the ordinary vacuum physics, have many new peculiarities:
there are no massless physical excitations-the gap of the order of $gT$
is always generated by the medium and all spectra are split-the new (or
earlier forbidden) spectrum branches arise which are well-separated, at
least, at finite momenta [1]. Today the most interesting results are
obtained for the fermionic collective excitations [2-6] which in the
medium demonstrate themselves very pronouncedly:  their new spectrum
branches - quasi-hole excitations are completely different from the
quasi-particle ones and they all are split even at zero momentum. The
properties of these new spectrum branches are very peculiar and a number
of them (such as the minimum of the quasi-hole branches at finite
momentum and the "wrong" relation between chirality and helicity) are
intensively discussed today to find new physical predictions [3,7]. At
the same time all these peculiarities being generated by the medium are
qualitatively the same for any hot and dense one: in particular, they
take place in the QED medium as well where the usual electron and
photon spectra are essentially modified. The electron spectrum in
the medium is split even at zero momentum and demonstrates four
different masses; the photon also has the finite mass but this mass is
the common limit both longitudinal and transverse spectrum branches.

The goal of this paper is to present explicitly the photon and
electron excitations in hot and dense QED using the Feynman gauge and
the one-loop self-energy. All spectra are split at finite momenta and
their branches develop the gap (the dynamical mass) at zero momentum.
For the case $m<<gT$ using the usual high temperature technique we find
collective spectra for all $|\q|$ and discuss their asymptotical
behaviors and gauge invariance. Everywhere the standard temperature
Green function technique is used and the case of zero damping is only
considered. In doing so many additional problems connected with the
gauge invariant calculations of damping [8] are excluded since they
being closely related to the infrared properties of an individual
theory are not the subject of this paper.

\section{ QED Lagrangian and photon self-energy}
The QED Lagrangian in covariant gauges has the form
\setcounter{equation}{0}
\begin{eqnarray}
{\cal L}=&-&\frac{1}{4}{F_{\mu\nu}}^2+{\bar \psi}[\gamma_{\mu}
(\partial_{\mu}-ie V_{\mu})+m]\psi \nonumber\\ &-&\mu{\bar
\psi}\gamma_4\psi +\frac{1}{2\alpha}(\partial_{\mu}V_{\mu})^2 +{\bar
C}\partial_{\mu}\partial_{\mu}C
\end{eqnarray}
where $F_{\mu\nu}=\partial_{\mu}V_{\nu}-\partial_{\nu}V_{\mu}$  is
the Abelian field strength; $\psi$(and ${\bar \psi}$) are the
Dirac fields  and $C$ (and ${\bar C}$) are the ghost Fermi fields.
In Eq.(1) $\mu$ and $m$ are the electron chemical potential and
the bare electron mass, respectively, and $\alpha$ is the gauge
fixing parameter ($\alpha=1$ for the Feynman gauge). The metrics is
chosen to be Euclidean, and $\gamma_{\mu}^2=1$.

\noindent
The exact Schwinger-Dyson equation for the photon Green function is
chosen to be
\begin{eqnarray}
{\cal D}^{-1}(k)={\cal D}_0^{-1}(k)+\Pi(k)
\end{eqnarray}
where the photon self-energy has the well-known representation
\begin{eqnarray}
-\Pi_{\mu\nu}(k)=\frac{e^2}{\beta}\sum_{p_4}^F
\int\frac{d^3p}{(2\pi)^3}\gamma_{\mu}G(p-k)
\Gamma_{\nu}(p-k,p|k)G(p)\,.
\end{eqnarray}
For hot QED it is also proved that the photon self-energy tensor is
transverse [9]
\begin{eqnarray}
k_{\mu}\Pi_{\mu\nu}(k)=0
\end{eqnarray}
and due to this fact it can be represented with the aid of two scalar
functions in the usual manner
\begin{eqnarray}
\Pi_{\mu\nu}(k_4,\k)=(\delta_{\mu\nu}-\frac{k_\mu k_\nu}{k^2})
A(k_4,\k)+(\frac{k_\mu k_\nu}{k^2}-\frac{k_\mu u_\nu+k_\nu u_\mu}{(uk)}
+\frac{u_\mu u_\nu }{(uk)^2}k^2)B(k_4,\k)
\end{eqnarray}
where $u_\mu$ is the unit medium vector.
The representation (5) leads to the photon propagator in the form [1]
\begin{eqnarray}
{\cal D}(k)&=&(\delta_{\mu\nu}-\frac{k_\mu k_\nu}{k^2})(k^2+A(k))^{-1}
+\frac{\alpha}{k^2}\frac{k_\mu k_\nu}{k^2}\nonumber\\
&+&\frac{B(k)\displaystyle{(\frac{k_\mu k_\nu}{k^2}-\frac{k_\mu
u_\nu+k_\nu u_\mu}{(uk)} +\frac{u_\mu u_\nu}
{(uk)^2}k^2)}}{[k^2+A(k)][k^2+A(k)+B(k)(k^2/(uk)-1)]}
\end{eqnarray}
where the gauge is fixed by the parameter $\alpha$, and its
longitudinal part, according to the Slavnov-Taylor identities, is not
changed by radiative corrections. According Eq.(6) there are two
dispersion relations
\begin{eqnarray}
k^2+A(k)=0   \,,\qquad    k^2+A(k)+B(k)(\frac{k^2}{(uk)}-1)=0
\end{eqnarray}
which for the rest system of the medium (where ${\bf u}=0$ and $u_4=1$)
have the following representation
\begin{eqnarray}
k_4^2+\k^2+A(k_4,\k)=0  \,,\qquad   \k^2+\Pi_{44}(k_4,\k)=0 \,.
\end{eqnarray}
Using the standard diagram technique it is enough to calculate for QED
(where Eq.(4) is reliably proved) only the $\Pi_{ij}(k_4,\k)$ tensor
and to restore the $A(k_4,\k)$ and $\Pi_{44}(k_4,\k)$ functions
following the exact relation
\begin{eqnarray}
\Pi_{ij}(k_4,\k)&=&(\delta_{ij}-\frac{k_ik_j}{\k^2})A(k_4,\k)+
\frac{k_ik_j}{\k^2}\frac{k_4^2}{\k^2}\Pi_{44}(k_4,\k)\,.
\end{eqnarray}
Of course the $\Pi_{44}(k_4,\k)$ function can be found independently that
sometimes is useful for checking the calculations made.

\section{ Collective Bose excitations in hot QED}
In the one-loop approximation one has
\begin{eqnarray}
-\Pi_{\mu\nu}(q)=\frac{e^2}{\beta}\sum_{p_4}^F\int\frac{d^3p}{(2\pi)^3}
\;\frac{\gamma_{\mu}[-i\gamma_{\rho}({\hat p}-q)_{\rho}+m]\gamma_{\nu}
[-i\gamma_{\sigma}{\hat p}_{\sigma}+m]}{[({\hat p}-q)^2+m^2]\;[{\hat
p}^2+m^2]}
\end{eqnarray}
where ${\hat p}=\{(p_4+i\mu),\p \}$
and standard calculations (at first the summation over
spinor indices and then over Fermi frequencies) should be performed to
obtain the final expressions for $\Pi_{44}(q)$ and $A(q)$ functions.
These expressions are [9]
\begin{eqnarray}
\Pi_{44}(q_4,\q)=\frac{e^2}{\pi^2}\int\frac{\p^2d|\p|}{\epsilon_\p}
(n_{\p}^++n_{\p}^-)\Bigr[\,1+\frac{\q^2+q_4^2-4\epsilon_\p^2}{8\p\q}
\ln{\bf a}-\frac{iq_4\epsilon_\p}{2\p\q}\ln{\bf b}\Bigr]
\end{eqnarray}
and
\begin{eqnarray}
\!\!\!\!\!\!\!\!\!\!\!\!\!\!\!
A(q_4,\q)&=&\frac{e^2}{2\pi^2}\int\frac{\p^2d|\p|}{\epsilon_\p}
(n_{\p}^++n_{\p}^-)\Bigr[\,1\nonumber\\
&-&\frac{q_4^2}{\q^2}
+\frac{\q^4-q_4^4+4\epsilon_\p^2\,q_4^2+4\p^2\q^2}{8\p\q^3}\ln{\bf a}
+\frac{iq_4\epsilon_\p}{2\p\q^3}(q_4^2+\q^2)\ln{\bf b}\Bigr]
\end{eqnarray}
where $\epsilon_\p=\sqrt{\p^2+m^2}$ is the bare electron energy
and $n_\p^{\pm}=\left\{\exp\beta(\;\epsilon_\p \pm \mu)+1\right\}^{-1}$
are the Fermi occupation numbers. The functions ${\bf a}$ and ${\bf b}$
which define logarithms have the form
\begin{eqnarray}
{\bf a}=\frac{(\q^2-2\p\q+q_4^2)^2+4q_4^2\epsilon_\p^2}
{(\q^2+2\p\q+q_4^2)^2+4q_4^2\epsilon_\p^2} \,,\qquad
{\bf b}=\frac{(\q^2+q_4^2)^2-4(\p\q+iq_4\epsilon_\p)^2}
{(\q^2+q_4^2)^2-4(\p\q-iq_4\epsilon_\p)^2}
\end{eqnarray}
and are rather typical for such calculations.

At first it is useful to find the long wavelength limits of both
longitudinal and transverse spectra remaining in the framework of
the one-loop approximation and making the standard analytical
continuation. In this case  $\Re \Pi_{44}(\omega,\q)$ and $\Re
A(\omega,\q)$ should be expended in powers of $|\q|$ (keeping
$\omega \ne 0$)
\begin{eqnarray}
-\Re \Pi_{44}(\omega,|\q|\rightarrow 0)=\frac{\q^2}{\omega^2}
(\omega_0^2+ b_L\frac{\q^2}{\omega^2})\,,\qquad \Re A(\omega,
|\q| \rightarrow 0)=\omega_0^2+b_T\frac{\q^2}{\omega^2}
\end{eqnarray}
and in this form they are used to solve Eq.(8).
Here new abbreviations are:
\begin{eqnarray}
\omega_0^2&=&\frac{e^2}{\pi^2}\int\frac{\p^2d|\p|}{\epsilon_\p}
(n_\p^++n_\p^-)(1-\frac{\p^2}{3\epsilon_\p^2})\,, \nonumber\\
b_L &=&\frac{e^2}{\pi^2}\int\frac{\p^2d|\p|}{\epsilon_\p}
(n_\p^++n_\p^-)\frac{\p^2}{\epsilon_\p^2}(1-\frac{3\p^2}{5\epsilon_\p^2})
\end{eqnarray}
and $ b_T=b_L/3 $ . Finally the long wavelength limits of the collective
photon excitations have the form
\begin{eqnarray}
\omega_L^2(\q)=\omega_0^2+\frac{b_L}{\omega_0^2}\q^2 \,,\qquad
\omega_T^2(\q)=\omega_0^2+(1+\frac{b_T}{\omega_0^2})\q^2
\end{eqnarray}
and demonstrate that these spectra have a gap and are split at finite
momenta.  This gap is a dynamical photon mass which is always nonzero
in the medium.

Now we continue to find the photon spectra for any momenta but only in
the high temperature limit considering that $T>>m,\mu$.  To this end we
replace $|\p|$ by $|\p|T$  inside all integrals within Eqs.(11) and
(12)  and simplify them keeping only the leading terms.  Then the
dispersion relations will be solved using these simplified integrals.

Our result when  $m\ne$ 0 is two rather complicated integral equations
which are separated for the longitudinal mode
\begin{eqnarray}
\omega_L^2(\q)=\xi^2\frac{e^2}{\pi^2}\int\frac{\p^2d|\p|}{\epsilon_\p}
(n_\p^++n_\p^-)\Bigr[\,\frac{\xi d_\p}{2}\ln\frac{(1+\xi d_\p)^2}
{(1-\xi d_\p)^2}-\frac{d_\p^2(1-\xi^2)}{1-\xi^2d_\p^2}-1\,\Bigr]
\end{eqnarray}
and for the transverse excitations
\begin{eqnarray}
\omega_T^2(\q)=\xi^2\frac{e^2}{\pi^2}\int\frac{\p^2d|\p|}{\epsilon_\p}
(n_\p^++n_\p^-)\Bigr[\,\frac{\xi^2}{\xi^2-1}-\frac{\xi
d_\p}{4}\ln\frac{(1+\xi d_\p)^2} {(1-\xi d_\p)^2}\,\Bigr] \,.
\end{eqnarray}
Here $ d_\p=\epsilon_\p/|\p| $ and the variable $\xi=iq_4/|/q|$ runs in
$1<\xi<\infty$ . The long wavelength limit corresponds to
$\xi\rightarrow\infty$.

To find these limits  one should expand in powers of $1/\xi\,$
Eq.(17) and (18) and then replace $\xi^2$ as $\xi^2 =\omega_0^2/\q^2$
where $\omega_0^2$ is the first term of these expansions. The result
has the form
\begin{eqnarray}
\omega_L^2(\q)=\frac{e^2}{\pi^2}\int\frac{\p^2d|\p|}{\epsilon_\p}
(n_\p^++n_\p^-)\Bigr[\,(1-\frac{1}{3d_\p^2})\,+\,\frac{1}{d_\p^2}\,
(1-\frac{3}{5d_\p^2})\,\frac{\q^2}{\omega_0^2} \,\Bigr]\,,
\end{eqnarray}
\begin{eqnarray}
\omega_T^2(\q)=\frac{e^2}{\pi^2}\int\frac{\p^2d|\p|}{\epsilon_\p}
(n_\p^++n_\p^-)\Bigr[\,(1-\frac{1}{3d_\p^2})\,+\,
(1-\frac{1}{5d_\p^4})\,\frac{\q^2}{\omega_0^2} \,\Bigr]
\end{eqnarray}
and reproduces the earlier found Eq.(16).
For the extreme high temperature when both $m/T<<1$ and $\mu/T<<1$ one
has
\begin{eqnarray}
\omega_L^2(\q)=\frac{e^2}{9\beta^2}+\frac{3}{5}\q^2 \,,\qquad
\omega_T^2(\q=\frac{e^2}{9\beta^2}+\frac{6}{5}\q^2 \,,\qquad
\end{eqnarray}
since this limit means that $m,\mu=0$  everywhere and all integrals are
easily calculated. The spectra (21) are gauge invariant, but if $m,\mu
\ne 0$ this invariance is very questionable.

In fact the real difficulty introduces only the
$m$-parameter and if $m=0$ one can easily simplify all
calculations. According to Eqs. (17) and (18) where now ${\bf d_\p}=1$,
the excitation spectra for any $|\q|$ have a rather simple form
\begin{eqnarray}
\omega_L^2(\q)\,=\,\frac{3}{2}\,\omega_0^2\,\xi^2\Bigr[\frac{\xi}{2}
\ln\frac{(\xi+1)^2}{(\xi-1)^2}-2\Bigr]\,,\qquad \omega_T^2\,
=\,\frac{3}{2}\,\omega_0^2\,\xi^2\,\Bigr[\frac{\xi^2}{\xi^2-1}-
\frac{\xi}{4}\ln\frac{(\xi+1)^2}{(\xi-1)^2}\,\Bigr]
\end{eqnarray}
and these expressions give a chance to investigate the photon
spectra in full detail. In Eq.(22)
\begin{eqnarray}
\omega_0^2&=&\frac{2e^2}{3\pi^2}\int |\p|\, d|\p|(n_\p^++n_\p^-)
\end{eqnarray}
that gives $\omega_0^2=e^2/9\beta^2$ if $\mu=0$ and in this case the
spectra (22) are gauge invariant for any $|\q|$.

In the high-momentum region the asymptotical behaviors found for the
transverse and the longitudinal excitations are completely
different. The transverse spectrum branches are approximated as
\begin{eqnarray}
\omega_T^2(q)&=&\q^2+\,\frac{3}{2}\,\omega_0^2
\end{eqnarray}
but this is not the case for the longitudinal mode where one finds a
more complicated expression
\begin{eqnarray}
\omega_L^2(q)&=&\q^2\Bigr[1+4\,\exp(-\frac{2\q^2}{3\omega_0^2}-2)\Bigr]\,.
\end{eqnarray}
Moreover the transverse excitations exist only in the region
$\omega>|\q|$  but the longitudinal mode overcomes line $\omega=|\q|$
and appears for $\omega<|\q|$.

\section{ Collective Fermi excitations in hot QED }

To find the collective Fermi excitations in hot QED we start with the
exact Schwinger-Dyson equation
\begin{eqnarray}
G^{-1}(q)=G_0^{-1}(q)+\Sigma(q)
\end{eqnarray}
and calculate the electron self-energy according to its exact
representation [9]
\begin{eqnarray}
\Sigma(q)=\frac{e^2}{\beta}\sum_{p_4}^F
\int\frac{d^3p}{(2\pi)^3}{\cal D}_{\mu\nu}(p-q)\gamma_{\mu}G(p)
\Gamma_{\nu}(p,q|p-q)\;.
\end{eqnarray}
Today $\Sigma(q)$ is reliably found only within the one-loop
approximation, where the bare Green functions are used to calculate
Eq.(27) and the Feynman gauge is usually fixed to make all calculations
more simple.  Within this approximation any results found for the
collective Fermi excitations in QED will be the same as the results
established earlier for QCD in [6], if the factor $(N^2-1)/2N$ is
omitted everywhere in these papers.  However we briefly repeat these
results to solve our task more completely.

After all calculations have been performed, the one-loop $\Sigma(q)$
in QED is given by
\begin{eqnarray}
&&\!\!\!\!\!\!\!\!\!\!\!\!\!\!\!\Sigma(q)=
-e^2\int\frac{d^3p}{(2\pi)^3}\;\left\{\;\Bigr[
\frac{1}{\epsilon_\p}\;\frac{n_\p^+\;[\gamma_4\epsilon_\p+(i\g\p+2m)]}
{[q_4+i(\mu+\epsilon_\p)\;]^2+(\q-\p)^2}\right.\nonumber\\
&&\!\!\!\!\!\!\!\!\!\!\!\!\!\!\!\!\!\! \left.+\;\frac{n_\p^B}{|\p|}
\;\frac{(|\p|+\mu-iq_4) \gamma_4-[i\g(\q-\p)+2m]}{[q_4+
i(\mu+|\p|)\;]^2+\epsilon_{\p-\q}^2}\;\Bigr]
-\Big[h.c.(m,\mu)\rightarrow-(m,\mu)\Big]\right\}
\end{eqnarray}
where $\epsilon_\p=\sqrt{\p^2+m^2}$ is the bare electron energy;
$n_\p^{B}=\left\{\exp\beta|\p|-1\right\}^{-1}$ and $n_\p^{\pm}=
\left\{\exp\beta(\;\epsilon_\p \pm \mu)+1\right\}^{-1}$ are the Bose
and Fermi occupation numbers, respectively.

\noindent
Further two new functions are introduced and one can rewrite the
expression (28) as follows
\begin {eqnarray}
\Sigma(q)=i\gamma_{\mu}K_{\mu}(q)+m\;Z(q)
\end{eqnarray}
that is necessary to find nonperturbatively the function G(q)
\begin{eqnarray}
G(q)=\frac{-i\gamma_{\mu}({\hat q_{\mu}}+K_{\mu})+m\;(1+Z)}
{({\hat q_{\mu}}+K_\mu)^2\;+\;m^2\;(1+Z)^2} \,.
\end{eqnarray}
The one-loop dispersion relation for the collective Fermi excitations
if $\alpha =1$ has the form
\begin {eqnarray}
[\;(iq_4-\mu)-{\bar K}_4]^2\;=\;\q^2\;(1+K)^2 +m^2(1+Z)^2
\end{eqnarray}
and, after the standard analytical continuation has been made, can be
solved analytically or numerically.

At first, the special interest is to find the long wavelength limit for
the collective Fermi excitations in the most general case (when $m,\mu
\ne 0$). To this end the perturbative calculations nearly the bare mass
shell are used and the result is
\begin{eqnarray} E(0)=\mu+\frac{1}{2}\Big[\eta\;m_R-I_B\Big]\pm
\sqrt{\;\frac{[\eta\;m_R-I_B]^2}{4} +(I_K+4\eta mI_B)} \,.
\end{eqnarray}
Eq.(32) at once demonstrates a number of very interesting properties. In
particular, the spectrum branches are split even at zero momentum and
demonstrate four well-separated effective masses:  two of them
are related to the quasi-particle excitations and the other two present
the new quasi-hole ones. This is evident from Eq.(32) where $\eta=\pm
1$.  Here $m_R=m(1-2I_Z)$ and new abbreviations are:
\begin {eqnarray}
&&I_K\;=\;e^2\int\limits_0^{\infty}\frac{d|\p|}{2\pi^2}
\;\Bigr[\;\epsilon_\p
\,\frac{n_\p^++n_\p^-}{2}\;+\;|\p|\,n_\p^B\;\Bigr] \,, \\
&&I_B=-e^2\int\limits_0^{\infty}
\frac{d|\p|}{4\pi^2}\;\frac{n_\p^+-n_\p^-}{2} \,,\qquad
I_Z=e^2\int\limits_0^{\infty}
\frac{d|\p|}{4\pi^2}\;\frac{n_\p^++n_\p^-}{2\epsilon_\p}\;.
\end{eqnarray}
Everywhere $T$ is arbitrary.

The spectra can be also investigated for any $|\q|$. But such
calculations are known only in the high temperature region and can be
performed within Eq.(31) if either $m$ or $\mu$ is equal to zero.

The case $m=0$ with $\mu\ne 0$ is more simple since in this case
everywhere $\mu$ combines with $E=ip_4$ and one can perform all
calculations in the standard manner. The final result has the form
\begin{eqnarray}
E(\xi)=\mu-\frac{\xi\;I_B}{2(\xi-\eta)}\;
\pm\sqrt{\frac{\xi^2I_B^2}{4(\xi-\eta)^2}+I_K\xi^2\Bigr(\;\frac{\eta}
{\xi-\eta}+\frac{\eta}{2}\ln\frac{\xi-1}{\xi+1}\;\Bigr)}
\end{eqnarray}
where the variable $\xi$ runs within the range $1<\xi<\infty$. The
long wavelength limit corresponds to $\xi\rightarrow\infty$ and, of
course, reproduces Eq.(32) if inside it $m=0$.

Another possibility when Eq.(31) can be solved exactly for all $|\q|$
is the case $m\ne 0$ but with $\mu=0$. This possibility is more
complicated and is sensitive to the ratio $m/gT$. But if $\mu=0$ a
number of simplifications always arise, and, keeping the accepted
accuracy of calculations, the spectra for the light $m$ have the form

\begin{eqnarray}
\omega_{\pm}(\xi)^2=\frac{\xi^2(\;2I_K+m_R^2)}{2(\xi^2-1)}\;
\pm\;\sqrt{\frac{\xi^4}{(\xi^2-1)^2}\Bigr[\;(b(\xi)I_K)^2
+\;m_R^2(\;I_K+m_R^2/4)\Bigr]}
\end{eqnarray}
and present the collective excitations of massive electrons in
hot medium for all $|\q|$ when $m<<gT$. Two spectrum branches (when
the sign in Eq.(36) is plus) correspond to the quasi-particle
excitations and the other two (when the sign is minus) are the
quasi-hole ones. These spectrum branches differ in their asymptotical
behavior and in many other properties (e.g. the minimum can be found
for the quasi-hole branches at the finite momentum).

The long-wavelength behavior of these spectra (when
$\xi\rightarrow\infty$) has the form
\begin {eqnarray}
\omega_{\pm}(|\q|)^2=\;M_{\pm}^2\;+\;\Bigr(\;M_{\pm}^2\pm
\frac{4}{9}\frac{I_K^2}{\sqrt{\;m_R^2(m_R^2+4I_K\;)}}
\;\Bigr)\;\frac{|\q|^2}{M_{\pm}^2}\;+\;O(|\q|^4)
\end{eqnarray}
where the effective masses squared are given by
\begin {eqnarray}
M_{\pm}^2=\;\frac{m_R^2}{2}+I_K\pm
\sqrt{\;m_R^2\Bigr(\frac{m_R^2}{4}+I_K\;\Bigr)}\;.
\end{eqnarray}
These masses which are the same as in Eq.(32) are different for four
spectrum branches $M_{\pm}=\frac{1}{2}(\eta m_R\pm \sqrt{m_R^2+4I_K})$
and always are nonzero in the medium. Here $\eta=\pm 1$. It is
also very important that the quasi-hole spectra $\omega_-(|\q|)^2$ are
very sensitive to the choice of the $m,T$-parameters.  In many cases
these spectra are the monotonous functions for small $|\q|^2$ and the
well-known minimum [2] disappears. This minimum always exists for the
massless particles, but when $m\ne 0$ the special conditions are
necessary to produce it.

In the high-momentum region the asymptotical behaviors found for the
quasi-particles and the quasi-hole excitations are completely
different. The quasi-particle spectrum branches are approximated as
follows
\begin{eqnarray} \omega_{+}(|\q|)^2=\;|\q|^2\;+\;(2I_K+m_R^2)\;
-\;\frac{I_K^2}{|\q|^2}\ln\frac{4|\q|^2}{2I_K+m_R^2}
\end{eqnarray}
where the nonanalytical term is unessential. Another situation takes
place for the quasi-hole excitations which do not exist in the vacuum
(when $T$ and $\mu$ are equal to zero). They very fast disappear and
their asymptotical behaviour is found to be
\begin{eqnarray}
\omega_{-}(|\q|)^2=
\;|\q|^2\;+4|\q|^2\exp(-\frac{|\q|^2(2I_K+m_R^2)}{I_K^2})\;.
\end{eqnarray}
In the high momentum region these spectrum branches approach the line
$\omega^2=|\q|^2$ more quickly than (39).

\section{Conclusion}
To summarize we have obtained the one-loop photon and electron spectra
for hot and dense QED.  All spectra are split and their branches
develop the gap (the dynamical mass) at zero momentum. These
dynamical masses are always nonzero in the medium and for many cases
are the gauge invariant quantities. In the case $m<<gT$ using the
high temperature technique the spectra are established for all $|\q|$
and their asymptotical behaviour is found explicitly. The photon
spectrum has two branches (longitudinal and transverse) which are
separated at all finite momenta but have the same limit (the dynamical
photon mass) at zero momentum. This is not the case for the collective
electron excitations which for any momentum (including zero one) are
split and in the general case (when $m,\mu \ne 0$) have four
well-separated branches:  two of them present the quasi-particle
excitations and two other correspond to the quasi-hole ones. The found
splitting demonstrates that the effective masses for all branches are
different when $m\ne 0$ and are always nonzero in the medium. The gauge
invariance of the results found in many cases is not proved and there
is no guarantee that this is indeed true, especially if the $m/T$ and
$\mu/T$- corrections are taken into account. This problem should be
investigated separately and most probably beyond the frame of one-loop
approximation since this approximation is very simple and does not make
any difference between the QED and QCD medium.  It is not ruled out that
the Braaten-Pisarski resummation is necessary to improve the situation,
but this question is not so evident as it is for the usual damping rate
calculations.

\begin{center}
{\bf References}
\end{center}

\renewcommand{\labelenumi}{\arabic{enumi}.)}
\begin{enumerate}

\item{ O.~K.~Kalashnikov and V.~V.~Klimov, Yad. Fiz. {\bf 31} (1980)
1357 (Soviet J. Nucl. Phys. {\bf 31} (1980) 699) and also the review
paper  O.~K.~Kalashnikov, Fortschr. Phys. {\bf 32} (1984) 525.}

\item{ V.~V.~Klimov, Yad.Fiz. {\bf 33} (1981) 1734 (Sov. J. Nucl. Phys.
{\bf 33} (1981) 934); Zh. Eksp. Teor. Fiz. {\bf 82} (1982) 336 (Sov.
Phys. JETP  {\bf 55} (1982) 199) and also H. Arthur Weldon, Phys. Rev.
D {\bf 26} (1982) 2789.}

\item{ R.~D.~Pisarski, Nucl. Phys. {\bf A498} (1989) 423c.}

\item{ G.~Baym, Jean-Paul Blaizot and B.~Svetitsky, Phys.Rev.
D {\bf 46}  (1992) 4043.}

\item{ C.~Quimbay and ~S.~Vargas-Castrillon, Nucl. Phys. {\bf B451}
(1995) 265.}

\item{ O.~K.~Kalashnikov, Mod. Phys. Lett. {\bf A12} (1997) 347 and
also  Pis'ma  Zh. Eksp. Teor. Fiz. {\bf 67} (1998) 3 (JETP Lett.
{\bf 67} (1998) No.1.}

\item{ H. Arthur Weldon, Phys. Rev. D {\bf 40} (1989) 2410.}

\item{ Jean-Paul Blaizot and E.Iancu, Phys. Rev. {\bf D55} (1997) 973
and also Preprint Saclay T97/052, hep-ph/9706397. }

\item{ E.~S.~Fradkin, Trudy Fiz. Inst. {\bf 29} (1965) 7 (Proc. of
P.~N.~Lebedev Physical Inst. {\bf 29} (1967) 1 ; translated into
English by Consultants Bureau, New York, 1967}

\end{enumerate}

\end{document}